\documentclass[preprint,amsmath,amssymb]{revtex4}
\usepackage{graphicx}
\usepackage{subfigure}
\usepackage[colorlinks,
            linkcolor=blue,
            anchorcolor=blue,
            citecolor=blue]{hyperref}
\newtheorem{theorem}{Theorem}

\def\ra{\rangle}
\def\la{\langle}
\allowdisplaybreaks[4]

\begin{document}

\title{Tighter Sum Uncertainty Relations via Variance and Wigner-Yanase Skew Information for $N$ Incompatible Observables}

\author{Qing-Hua Zhang$^{1,}$\footnotemark[1]}
\author{Shao-Ming Fei$^{1,2,}$\footnotemark[1]}

\affiliation{$^1$School of Mathematical Sciences, Capital Normal University,
Beijing 100048, China\\
$^2$Max-Planck-Institute for Mathematics in the Sciences, 04103 Leipzig, Germany}

\renewcommand{\thefootnote}{\fnsymbol{footnote}}

\footnotetext[1]{Corresponding authors. \\
Email address: \href{mailto:2190501022@cnu.edu.cn}{2190501022@cnu.edu.cn}.\\
Email address: \href{mailto:feishm@cnu.edu.cn}{feishm@cnu.edu.cn}.}

\bigskip

\begin{abstract}
We study the sum uncertainty relations based on variance and skew information for arbitrary finite $N$ quantum mechanical observables. We derive new uncertainty inequalities which improve the exiting results about the related uncertainty relations. Detailed examples are provided to illustrate the advantages of our uncertainty inequalities.
\end{abstract}

\maketitle
\section{INTRODUCTION}
As one of the fundamental building blocks of quantum theory, the uncertainty principle has attracted considerable attention since the innovation of quantum mechanics. Ever since Heisenberg proposed various notions of uncertainties related to the measurements of non-commuting observables in 1927 \cite{h1927}, a lot of researches have been done on quantifying the uncertainty of measurement outcomes, such as in terms of the noise and disturbance \cite{om2004,blw2013}, successive measurements  \cite{dd1983,smd2003,dp2013,bfs2014,zzy2015,cf2015}, informational recourses \cite{ww2010}, entropic terms  \cite{mu1988,pmms2017,fss2020,wym2009,r2013,rp2014,npg2016}, Wigner-Yanase skew information \cite{ls2003, cfl2016, zgy2021} and majorization techniques \cite{br2011,prz2013,fgg2013} etc..

Based on the variance of two arbitrary observables $A$ and $B$, Robertson derived the following well-known uncertainty relation \cite{r1929},
\begin{equation}\label{a1}
\Delta A\Delta B\geq \frac{1}{2}|\la \psi [A,B]\psi\ra|
\end{equation}
where $\Delta \Omega=\sqrt{\la \Omega^2\ra-\la\Omega\ra^2}$ is the standard deviation of an observable $\Omega$ and  the commutator $[A,B]=AB-BA$. For measurements on suitable states, the above uncertainty relation is nontrivial for non-commuting observables, namely, the non-commutativity of two arbitrary observables could be captured by the non-zero lower bound in (\ref{a1}). However, when the measured state $|\psi\ra$ is an eigenvector of either $A$ or $B$, lower bound in (\ref{a1}) is trivially zero. To deal with such problems, uncertainty relations based on the sum of variances have been taken into account. In \cite{mp2014} Maccone and Pati presented the following relations,
\begin{equation}\label{a2}
(\Delta A)^2+(\Delta B)^2\geq \pm\la\psi|[A,B]|\psi\ra + |\la\psi |A\pm iB|\psi^\bot \ra|^2,
\end{equation}
\begin{equation}\label{a3}
(\Delta A)^2+(\Delta B)^2\geq \frac{1}{2}[\Delta(A+B)]^2,
\end{equation}
where $\la \psi|\psi^\bot\ra=0$ and the signs $\pm$ on the right-hand side of (\ref{a2}) are so taken such that the lower bound attains the maximum.

Besides the uncertainty relations related to pairs of incompatible observables like position and momentum, the uncertainty relations with respect to three incompatible observables like the three components of spins and angular momentums \cite{kw2014,zdsw2015,sq2016,wbym2017} have been also investigated. Uncertainty relations for general multiple observables have been further studied either in product form \cite{qfl2016,xj2016} or sum form of variances \cite{ccfl2016, cbf2015, slpq2017, ccf2016, cwl2019}.
Song $et\ al.$ derived in \cite{slpq2017} an improved variance-based uncertainty relation,
\begin{equation} \label{a5}
\sum_{i=1}^N(\Delta A_i)^2 \geq \frac{1}{N}\Bigg\{ [\Delta (\sum_{{i=1}}^{N}  A_i)]^2+\frac{2}{N(N-1)}[\sum_{1\leq i<j\leq N} \Delta (A_i-A_j)]^2 \Bigg\}
\end{equation}
for arbitrary $N$ incompatible observables, which is stronger than the one derived from the uncertainty inequality for two observables \cite{mp2014}.

The skew information also provides a way to characterize uncertainty relation \cite{ls2003}. The Wigner-Yanase skew information of a state $\rho$ with respect to an operator $A$ is given by \cite{wy1963},
\begin{equation}
I_{\rho}(A)=-\frac{1}{2}tr([\sqrt{\rho},A]^2)=\frac{1}{2}\| [\sqrt{\rho},A]\|^2,
\end{equation}
where $\|\bullet \|$ denotes the Frobenius norm. The Wigner-Yanase skew information characterizes the intrinsic features of the state $\rho$ and the observable $A$. It is the same as the variance for pure states, but generally fundamentally different from the variance \cite{lz2004}. The skew information describes the non-commutativity between the square root of $\rho$ and the observable, while the variance describes the non-commutativity between the state $\rho$ and the observable.

In ref. \cite{cfl2016}, Chen $et\ al.$ provided the sum uncertainty relation based on Wigner-Yanase skew information for finite $N$ observables,
\begin{equation}\label{a6}
\sum_{i=1}^N I_{\rho}(A_i) \geq \frac{1}{N-2}\Bigg\{ \sum_{1\leq i<j\leq N} I_{\rho}(A_i+A_j)-\frac{1}{(N-1)^2}[\sum_{1\leq i<j\leq N} \sqrt{I_{\rho}(A_i+A_j)}]^2 \Bigg\}.
\end{equation}
Recently, Zhang $et\ al.$ improved the above sum uncertainty relation \cite{zgy2021},
\begin{equation} \label{a7}
\sum_{i=1}^N I_{\rho}(A_i) \geq \frac{1}{N}\Bigg\{ I_{\rho}(\sum_{{i=1}}^{N}  A_i)+\frac{2}{N(N-1)}[\sum_{1\leq i<j\leq N} \sqrt{I_{\rho}(A_i-A_j)}]^2 \Bigg\}.
\end{equation}
Both uncertainty inequalities (\ref{a6}) and (\ref{a7}) capture the incompatibility of the observables in the sense that their lower bounds are nonzero as long as the observables are not commutative with the measured state.

In this paper, we investigate sum uncertainty relations based on Wigner-Yanase skew information and variance for arbitrary $N$ incompatible observables. In Sec. \uppercase\expandafter{\romannumeral2}, we present a pair of uncertainty relation inequalities in term of variance, and we compare the uncertainty relation with existing ones for detail example, which shows our uncertainty relations can provide tighter bounds than others. In Sec. \uppercase\expandafter{\romannumeral3}, we obtain two uncertainty relations via skew information. And detail examples show the validity and superiority of our theorem to capture incompatibility. Then we conclude in Sec.\uppercase\expandafter{\romannumeral4}.

\section{UNCERTAINTY RELATION VIA VARIANCE}

In this section, we study stronger sum uncertainty relations based on the eigensystems of the observables. Let $A_i=\sum_k^du_{ik}| u_{ik}\ra\la u_{ik}|$ be the observable with the $k$-th eigenvalue $u_{ik}$ and the eigenstate $|u_{ik}\ra$. Then the variance is given by $(\Delta A_i)^2=\sum_k^d \bar{u}_{ik}^2\la| u_{ik}\ra\la u_{ik}|\ra$, where $\bar{u}_{ik}=u_{ik}-\la A_i\ra$ and $\la| u_{ik}\ra\la u_{ik}|\ra$ is the projective probability of
measured state in the basis $|u_{ik}\ra$. Set $a_i=(a_{i1},a_{i2},\dots,a_{id})=(|\bar{u}_{i1}|\sqrt{\la |u_{i1}\ra\la u_{i1}|\ra},|\bar{u}_{i2}|\sqrt{\la |u_{i2}\ra\la u_{i2}|\ra},\dots,|\bar{u}_{d1}|\sqrt{\la |u_{id}\ra\la u_{id}|\ra})$. Then $(\Delta A_i)^2=\sum_k^d a_{ik}^2=\|a_i\|^2$.
We have the following conclusion.

\begin{theorem}
Let  $A_1, A_2, \dots, A_N$ be arbitrary $N$ observables. The following variance-based sum uncertainty relation holds for any quantum state $\rho$,
\begin{equation} \label{th3eq1}
\sum_{i=1}^N(\Delta A_i)^2 \geq \max_{\pi_i,\pi_j \in S_d} \frac{1}{2N-2}\Bigg\{ \sum_{1\leq i<j\leq N} \Lambda_{\pi_i(i)\pi_j(j)}^2+\frac{2}{N(N-1)}[\sum_{1\leq i<j\leq N} \bar{\Lambda}_{\pi_i(i)\pi_j(j)}]^2 \Bigg\},
\end{equation}
where
\[\begin{aligned}&{\Lambda }_{\pi_i(i)\pi_j(j)}^2=\sum_{k=1}^d (|\bar{u}_{i{\pi_i(k)}}|\sqrt{\langle
|u_{i{\pi_i(k)}}\rangle \langle u_{i{\pi_i(k)}}|\rangle }+|\bar{u}_{j{\pi_j(k)}}|\sqrt{\langle
|u_{j{\pi_j(k)}}\rangle \langle u_{j{\pi_j(k)}}|\rangle })^2,\\&\bar{\Lambda
}_{\pi_i(i)\pi_j(j)}^2=\sum_{k=1}^d (|\bar{u}_{i{\pi_i(k)}}|\sqrt{\langle |u_{i{\pi_i(k)}}\rangle \langle
u_{i{\pi_i(k)}}|\rangle }-|\bar{u}_{j{\pi_j(k)}}|\sqrt{\langle |u_{j{\pi_j(k)}}\rangle \langle
u_{j{\pi_j(k)}}|\rangle })^2, \end{aligned}\]
$\pi_i,\pi_j\in S_d$ are arbitrary $d$-element permutation.
\end{theorem}

{\sf [Proof]} To prove the inequality (\ref{th3eq1}), we need the following identity for vectors $a_i$,
\begin{equation*}
(2N-2)\sum_{i=1}^{N} \| a_i\|^2 = \sum_{1\leq i<j \leq N} \| a_i+a_j \|^2 + \sum_{1\leq i<j \leq N} \| a_i-a_j \|^2,
\end{equation*}
where  $\|\bullet \|$ stands for the norm of a vector defined by inner product. Using the Cauchy-Schwarz inequality,
\begin{equation*}
\sum_{1\leq i<j \leq N} \| a_i - a_j \|^2 \geq \frac{2}{N(N-1)} (\sum_{1\leq i<j \leq N} \| a_i - a_j \|)^2,
\end{equation*}
we obtain
\begin{equation}\label{th3pf1}
\sum_{i=1}^{N} \| a_i^{\pi_i} \|^2 \geq \frac{1}{2N-2}[\frac{2}{N(N-1)}(\sum_{1\leq i<j \leq N} \| a_i^{\pi_i} - a_j^{\pi_j}  \|)^2 + \sum_{1\leq i<j \leq N} \| a_i ^{\pi_i} + a_j^{\pi_j}  \|^2],
\end{equation}
where \(a_i^{\pi_i}=(a_{i{\pi_i(1)}},a_{i{\pi_i(2)}},\dots ,a_{i{\pi_i(d)}}).\) This completes the proof. $\Box$

Arranging the components of the vector $b_i^{\uparrow}=(b_{i1},b_{i2},\dots, b_{id})^{\uparrow}=(|\bar{u}_{i1}|\sqrt{\la |u_{i1}\ra\la u_{i1}|\ra},|\bar{u}_{i2}|\sqrt{\la |u_{i2}\ra\la u_{i2}|\ra},\dots,|\bar{u}_{id}|\sqrt{\la |u_{id}\ra\la u_{id}|\ra})^{\uparrow}$
in increasing order, that is, $b_{ik} \leq b_{ik+1}$,
Chen $et\ al.$ introduced in \cite{cwl2019} a stronger uncertainty relation,
\begin{equation} \label{chen1}
\sum_{i=1}^N(\Delta A_i)^2 \geq \frac{1}{2^{H(2-N)}N-2}\Bigg\{ \sum_{1\leq i<j\leq N} K_{ij}^2+\frac{H(2-N)-1}{(N-1)^2}(\sum_{1\leq i<j\leq N} K_{ij})^2 \Bigg\},
\end{equation}
where $$K_{ij}^2= \sum_{k=1}^d (b_{ik}^{\uparrow} + b_{jk}^{\uparrow})^2,$$
and $H(x)$ is the unit step function with zero for $x<0$ and one for $x\geq 0$.

For two incompatible observables case, the inequality (\ref{th3eq1}) becomes
\begin{equation*}
(\Delta A_1)^2+(\Delta A_2)^2 \geq \max_{\pi_1,\pi_2 \in S_2} \frac{1}{2}\Bigg\{ \Lambda_{\pi_1(1)\pi_2(2)}^2+ \bar{\Lambda}_{\pi_1(1)\pi_2(2)}^2 \Bigg\}.
\end{equation*}
The uncertainty relation (\ref{chen1}) gives rise to
\begin{equation*}
(\Delta A_1)^2+(\Delta A_2)^2 \geq \frac{1}{2}K_{12}^2.
\end{equation*}
Since $\max_{\pi_1,\pi_2 \in S_2} \Lambda_{\pi_1(1)\pi_2(2)}^2=K_{12}^2$,
our Theorem 1 has tighter lower bound than (\ref{chen1}) due to an extra non-negative term.

\begin{figure}[htb]
 \centering
 \includegraphics[width=15cm]{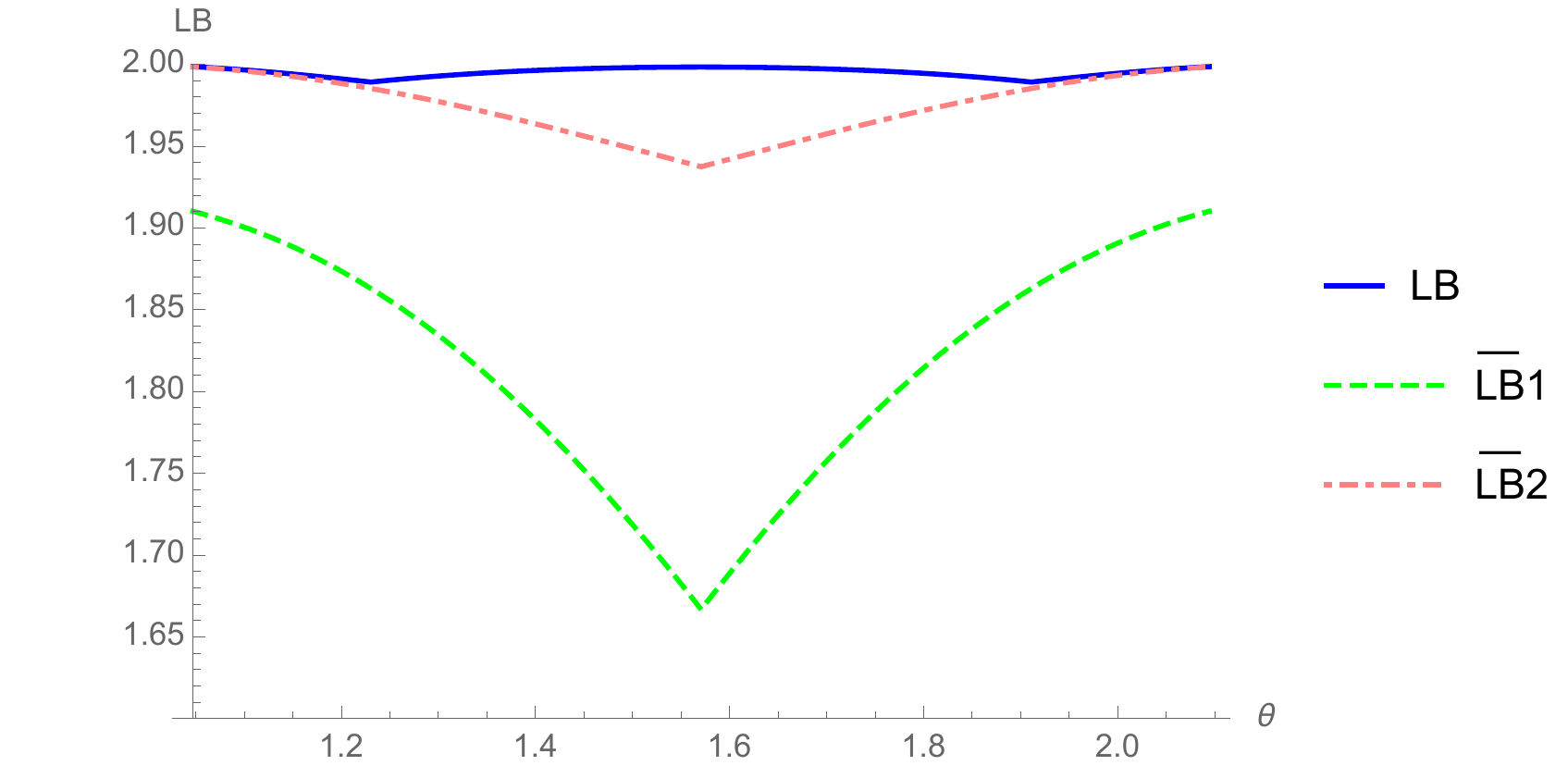}
 \caption{ Blue (solid), pink (dot-dashed), Green (dashed) and line represent the lower bounds of (\ref{th3eq1}), (\ref{chen1}) and (\ref{a5}), respectively.}
 \label{fig1}
 \end{figure}

\emph{Example 1.}
To illustrate that our uncertainty inequality (\ref{th3eq1}) is tighter than (\ref{a5}) and (\ref{chen1}), we consider the pure state,
\begin{equation}\label{ex1}
|\psi\ra =cos{\frac{\theta}{2}}|1\ra+e^{i\phi}sin{\frac{\theta}{2}}|0\ra,
\end{equation}
where $0\leq\theta\leq \pi$ and $0\leq\phi\leq 2\pi$. We take observables $A_1=-|0\ra\la1|-|1\ra\la0|$, $A_2=-i|0\ra\la1|+i|1\ra\la0|$ and $A_3=|0\ra\la0|-|1\ra\la1|$.
Set $\phi=\pi/4$. The comparison between the lower bounds from (\ref{th3eq1}), (\ref{a5}) and (\ref{chen1}) is shown in Fig. \ref{fig1}. For the sake of simplicity, ${\rm LB}$, $\overline{\rm LB}1$ and $\overline{\rm LB}2$  represent the lower bounds of (\ref{th3eq1}), (\ref{a5}) and (\ref{chen1}), respectively. The bound of (\ref{th3eq1}) is tighter than (\ref{a5}) and (\ref{chen1}) for certain $\theta$.

\section{UNCERTAINTY RELATIONS VIA SKEW INFORMATION}

We now provide stronger sum uncertainty relation inequalities based on
Wigner-Yanase skew information for $N$-incompatible observables.

\begin{theorem}\label{th2}
For arbitrary finite $N$ observables $A_1, A_2, \dots, A_N$, the following  sum uncertainty relations hold:
\begin{equation}\label{th2eq1}
\sum_{i=1}^N I_{\rho} (A_i)\geq \frac{1}{2N-2}\Bigg\{\frac{2}{N(N-1)} [\sum_{1\leq i<j\leq N} \sqrt{I_{\rho}(A_i+A_j)}]^2+\sum_{1\leq i<j\leq N} I_{\rho} (A_i-A_j) \Bigg\},
\end{equation}
and
\begin{equation}\label{th2eq2}
\sum_{i=1}^N I_{\rho} (A_i)\geq \frac{1}{2N-2}\Bigg\{ \frac{2}{N(N-1)}[\sum_{1\leq i<j\leq N} \sqrt{I_{\rho} (A_i-A_j)}]^2 + \sum_{1\leq i<j\leq N} I_{\rho}(A_i+A_j)  \Bigg\}.
\end{equation}
\end{theorem}

{\sf [Proof]}
Using the following identity for any Hermitian matrices $a_i$,
\begin{equation*}
(2N-2)\sum_{i=1}^{N} \| a_i\|^2 = \sum_{1\leq i<j \leq N} \| a_i+a_j \|^2 + \sum_{1\leq i<j \leq N} \| a_i-a_j \|^2,
\end{equation*}
where  $\|\bullet \|$ stands for the Frobenius norm,
and the Cauchy-Schwarz inequalities,
\begin{equation*}
\sum_{1\leq i<j \leq N} \| a_i + a_j \|^2 \geq \frac{2}{N(N-1)}
(\sum_{1\leq i<j \leq N} \| a_i - a_j \|)^2,
\end{equation*}
and
\begin{equation*}
\sum_{1\leq i<j \leq N} \| a_i - a_j \|^2 \geq \frac{2}{N(N-1)} (\sum_{1\leq i<j \leq N} \| a_i - a_j \|)^2,
\end{equation*}
we have
\begin{equation}\label{th2pf1}
\sum_{i=1}^{N} \| a_i\|^2 \geq \frac{1}{2N-2}[\frac{2}{N(N-1)}(\sum_{1\leq i<j \leq N} \| a_i + a_j \|)^2 + \sum_{1\leq i<j \leq N} \| a_i - a_j \|^2]
\end{equation}
and
\begin{equation}\label{th2pf2}
\sum_{i=1}^{N} \| a_i\|^2 \geq \frac{1}{2N-2}[\frac{2}{N(N-1)}(\sum_{1\leq i<j \leq N} \| a_i - a_j \|)^2 + \sum_{1\leq i<j \leq N} \| a_i + a_j \|^2].
\end{equation}
Denote $a_i=[\sqrt{\rho},A_i]$. Then $\| a_i \|^2=2I_{\rho} (A_i)$,
$\| a_i - a_j \|^2=2I_{\rho} (A_i - A_j)$ and $\| a_i + a_j \|^2=2I_{\rho} (A_i + A_j)$.
Substituting the above relations into the inequalities (\ref{th2pf1}) and (\ref{th2pf2})
we obtain (\ref{th2eq1}) and (\ref{th2eq2}). $\Box$

In fact, by using the parallelogram law in Hilbert space,
$I_{\rho} (A)+I_{\rho} (B) \geq \frac{1} {2}I_{\rho}(A+B)$ and
$I_{\rho} (A)+I_{\rho} (B) \geq \frac{1} {2}I_{\rho}(A-B)$, one can also get for $N$ observables,
\begin{equation}\label{a10}
\sum_{i=1}^N I_{\rho} (A_i)\geq \frac{1}{2N-2} \sum_{1\leq i<j\leq N} I_{\rho}(A_i+A_j)
\end{equation}
and
\begin{equation}\label{a11}
\sum_{i=1}^N I_{\rho} (A_i)\geq \frac{1}{2N-2} \sum_{1\leq i<j\leq N} I_{\rho}(A_i-A_j).
\end{equation}
Obviously, our Theorem 2 provides tighter uncertainty inequalities than the above inequalities (\ref{a10}) and (\ref{a11}) due to the extra non-negative terms.

We give two examples to show that our inequalities are tighter than (\ref{a7}). For convenience, $\overline{\rm Lb}$, Lb1 and Lb2 are employed to represent the right hands of  (\ref{a7}), (\ref{th2eq1}) and (\ref{th2eq2}), respectively.

\begin{figure}[tbp]
 \centering
 \includegraphics[width=15cm]{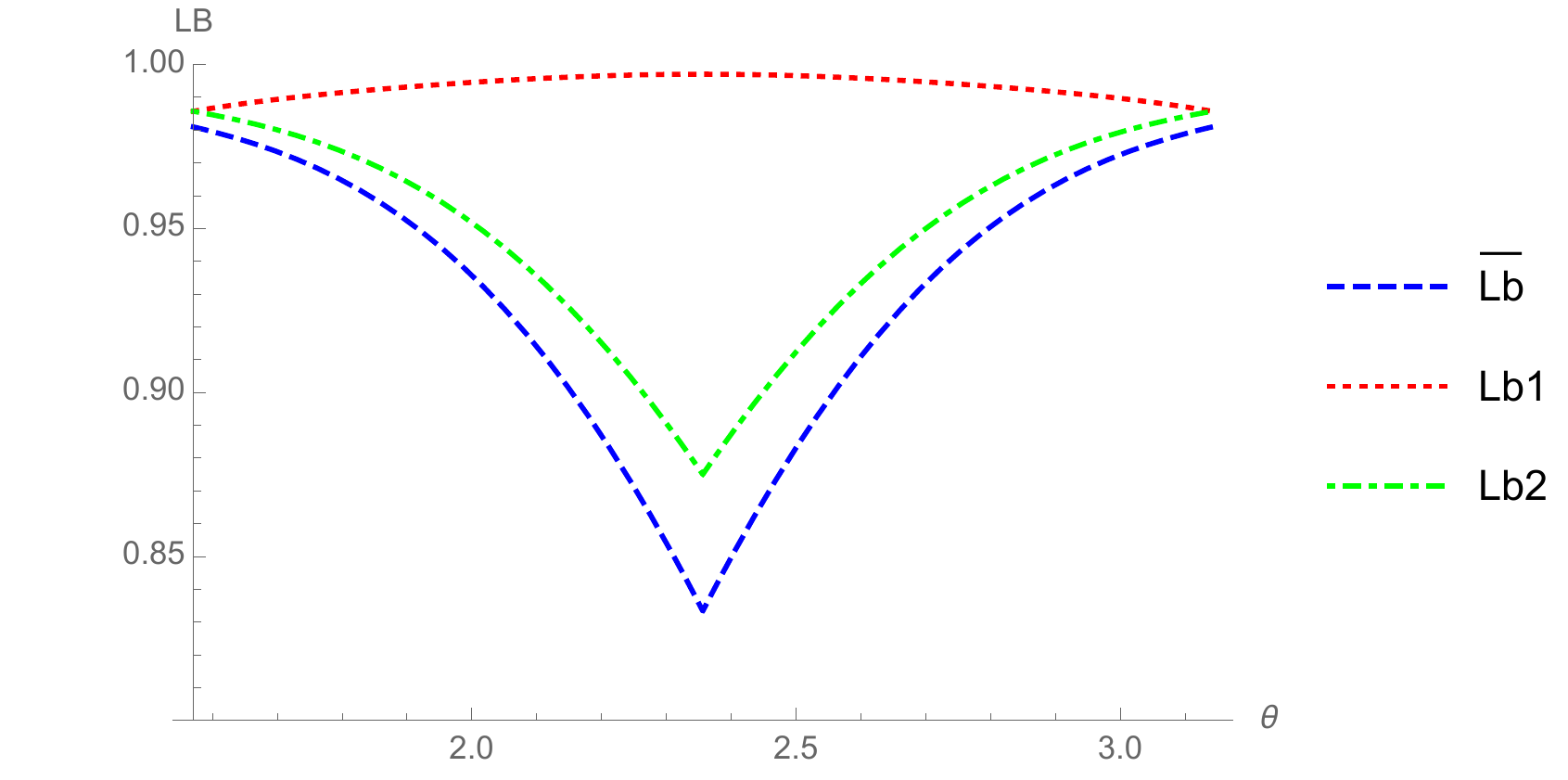}
 \caption{Blue (dashed) line is the bound (\ref{a7}). Red (dotted) and green (dot-dashed) lines represent the bounds of (\ref{th2eq1}) and (\ref{th2eq2}), respectively. Clearly, the bound of (\ref{th2eq1}) and (\ref{th2eq2}) are strictly tighter than the bound of (\ref{a7}) for some cases.}
  \label{fig2}
 \end{figure}

\emph{Example 2.}
We first consider the mixed state given by Bloch vectors $\vec{r}=\{ \frac{\sqrt{3}}{2}cos{\theta}, \frac{\sqrt{3}}{2}sin{\theta}, 0 \}$,
\begin{equation}
\rho=\frac{I_2 + \vec{r}\cdot\vec{\sigma}}{2},
\end{equation}
where $\vec{\sigma} = \{\sigma_x, \sigma_y, \sigma_z\}$ is a vector given by standard Pauli matrices and $I_2$ is the $2\times2$ identity matrix.

Choosing $\sigma_x$, $\sigma_y$ and $\sigma_z$ as the observables, we get
$$I_{\rho}(\sigma_x)+I_{\rho}(\sigma_y)+I_{\rho}(\sigma_z) = 1,\quad I_{\rho}(\sigma_x + \sigma_y+ \sigma_z)= 1-cos\theta sin\theta,$$
$$I_{\rho}(\sigma_x + \sigma_y) = \frac{1}{2}-cos\theta sin\theta, \quad I_{\rho}(\sigma_x + \sigma_z) =\frac{1}{4}(3-cos2\theta),\quad I_{\rho}(\sigma_y + \sigma_z) = \frac{1}{4}(3+cos2\theta), $$
$$ I_{\rho}(\sigma_x - \sigma_y) = \frac{1}{2}(1+sin2\theta), \quad I_{\rho}(\sigma_x - \sigma_z) =\frac{1}{4}(3-cos2\theta),\quad I_{\rho}(\sigma_y - \sigma_z) =\frac{1}{4}(3+cos2\theta).$$
The comparison between the lower bounds of Theorem 2 and (\ref{a7}) is given in Fig. \ref{fig2}.

\begin{figure}[tbp]
 \centering
 \subfigure[]
 {
 \label{fig:subfig:a} 
 \includegraphics[width=8cm]{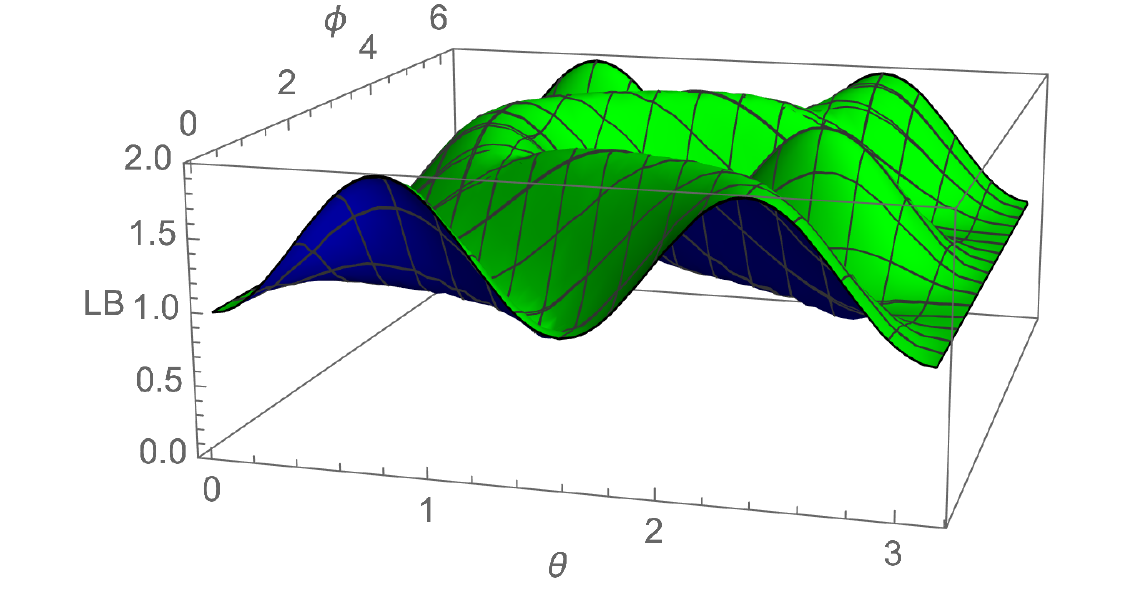}
 }
 \subfigure[]
 {
 \label{fig:subfig:b} 
 \includegraphics[width=7.5cm]{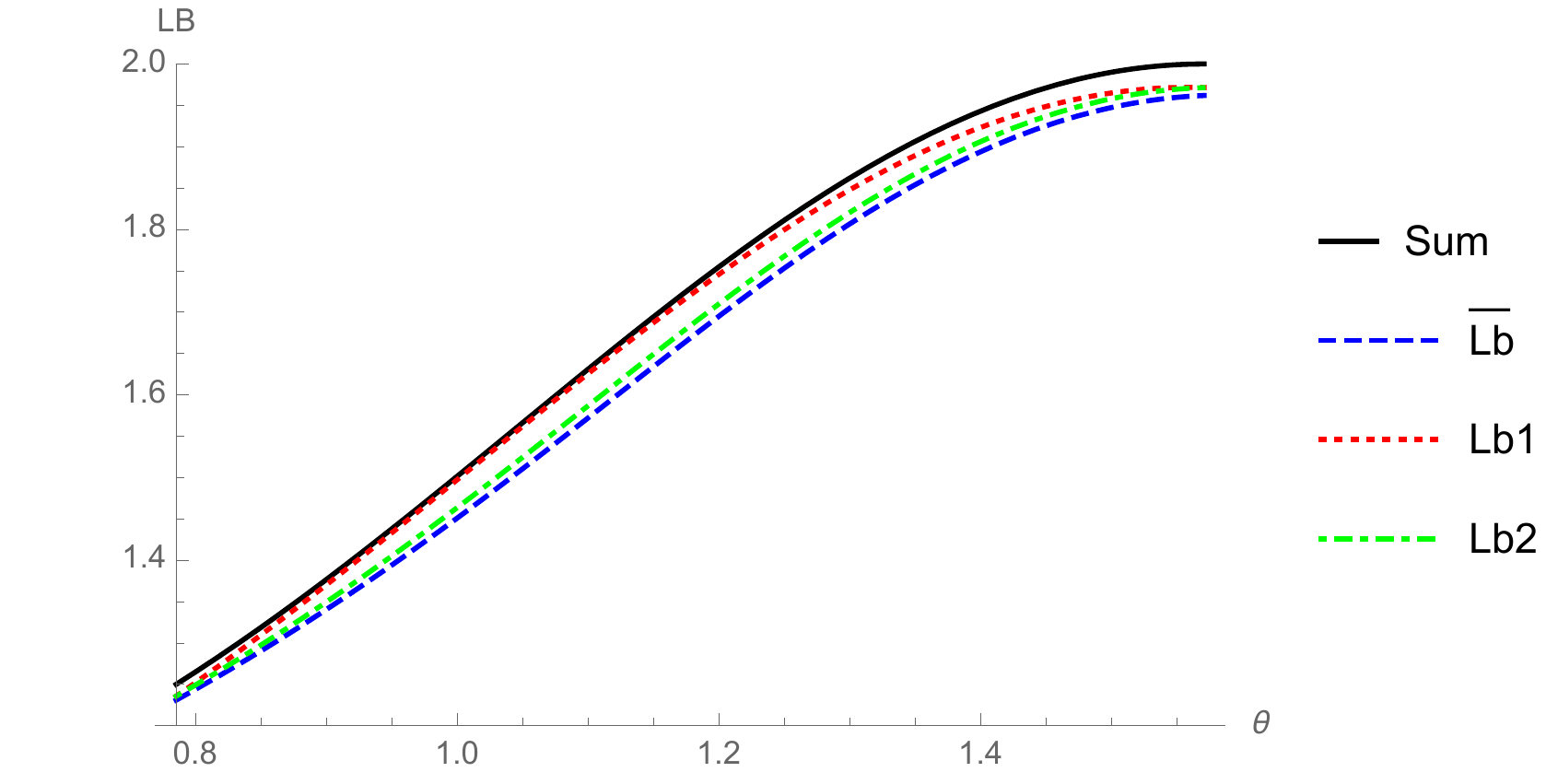}}
 \caption{(\textbf{a}) The green and blue surfaces represent our lower bound (\ref{th2eq2}) and the lower bound (\ref{a7}), respectively. Obviously, the green surface covers all the blue one. (\textbf{b}) Set $\phi=\pi/2$. Black (solid) line represents the sum of skew information $I_{\rho} (L_x)+I_{\rho}(L_x)+I_{\rho}(L_y)$. Blue (dashed) is the bound (\ref{a7}), red (dotted) and green (dot-dashed) lines represent the bounds of (\ref{th2eq1}) and (\ref{th2eq2}), respectively. The lower bounds (\ref{th2eq1}) and (\ref{th2eq2}) in Theorem 2 are tighter than (\ref{a7}) for certain cases.}
 \label{fig3}
 \end{figure}

\emph{Example 3.} We consider the following quantum state in spin-$1$ system,
\begin{equation}
|\psi\ra = sin{\theta}cos{\phi}|1\ra + sin{\theta}sin{\phi}|0\ra + cos{\theta}|-1\ra ,
\end{equation}
where $\theta\in [0, \pi]$ and $\phi\in [0,2\pi]$. We take angular momentum operators ($\hbar=1$) as the observables:
\begin{equation*}
\begin{gathered}
L_x=\frac{1}{\sqrt{2}}
\begin{pmatrix} 0 & 1 & 0 \\ 1 & 0 & 1 \\ 0 & 1&0 \end{pmatrix},
\quad
L_y=\frac{1}{\sqrt{2}}
\begin{pmatrix} 0 & -i & 0 \\ i & 0 & -i \\ 0 & i&0 \end{pmatrix},
\quad
L_z=
\begin{pmatrix} 1 & 0 & 0 \\ 0 & 0 & 0 \\ 0 & 0&-1 \end{pmatrix}.
\end{gathered}
\end{equation*}
We have the sum of skew information for the state $|\psi\ra$
\begin{equation*}
{\rm Sum}:=I_{\rho} (L_x)+I_{\rho}(L_x)+I_{\rho}(L_y)=2-(cos^2\theta -sin^2\theta cos^2\phi)^2-2sin^2\theta sin^2\phi(cos\theta+sin\theta cos\phi)^2.
\end{equation*}
We show in Fig. \ref{fig3} the comparison between the lower bounds (\ref{th2eq1}), (\ref{th2eq2}) and (\ref{a7}). The figures show that our bounds are better in this case.

\section{CONCLUSION}
We have derived uncertainty inequalities for arbitrary $N$-incompatible observables based on the sum of variance and Wigner-Yanase skew information, which improve the existing results about the related uncertainty inequalities. The simple approach used in this work can be also applied to investigate other kind of uncertainty relations.

\bigskip
\noindent{\bf Acknowledgments}\, This work is supported by NSFC (Grant No. 12075159), Beijing Natural Science Foundation (Z190005), Academy for Multidisciplinary Studies, Capital Normal University, the Academician Innovation Platform of Hainan Province, Shenzhen Institute for Quantum Science and Engineering, Southern University of Science and Technology (No. SIQSE202001).

\end{document}